\documentclass[%
 aip,
 amsmath,amssymb,
 reprint,%
]{revtex4-1}

\usepackage{graphicx}
\usepackage{dcolumn}
\usepackage{bm}

\usepackage[utf8]{inputenc}
\usepackage[T1]{fontenc}
\usepackage{etoolbox}
\usepackage{xcolor}

\begin{document}

\preprint{AIP/123-QED}

\title{Designing active colloidal folders}

\author{S. Das}
\affiliation{%
Department of Chemistry, Columbia University\\ 3000 Broadway, New York, NY 10027\\
}%
\affiliation{%
Department of Polymer Science and Engineering, University of Massachusetts Amherst\\
120 Governors Drive, Amherst, MA 01003
}%
\author{M. Lange}
\author{A. Cacciuto}
\email{ac2822@columbia.edu}
\affiliation{%
Department of Chemistry, Columbia University\\ 3000 Broadway, New York, NY 10027\\
}%

\date{\today}

\begin{abstract}
\noindent 
Can active forces be exploited to drive the consistent collapse of an active polymer into a folded structure? In this paper we introduce and perform numerical simulations of a simple model of active colloidal folders, and show that a judicious inclusion of active forces into a stiff colloidal chain can generate designable and reconfigurable two dimensional folded structures. The key feature is to organize the forces perpendicular to the chain backbone according to specific patterns (sequences).
We characterize the physical properties of this model and perform, using a number of numerical techniques,  an in-depth statistical analysis of structure and dynamics of the emerging conformations. We discovered a number of interesting features, including the existence of a 
direct correspondence between the sequence of the active forces and the structure of folded conformations, and we discover the existence of an ensemble of highly mobile compact structures capable of moving from conformation to conformation. Finally, akin to protein design problems, we discuss a method that is capable of designing specific target folds by sampling over sequences of active forces. 
\end{abstract}

\maketitle
 
\section{Introduction}

One of the most exciting problems in materials engineering is that of structure design via self-assembly. The problem of how to predict the final arrangement of a number of particles  into target conformations by manipulating their pairwise interactions is a hard one. Progress in this direction has been made with DNA origami and with  DNA-coated colloidal particles ( see~\cite{Dey2021Jan} and~\cite{Zhang2021} for recent reviews on these two topics respectively). Over the past decade, a new type of colloidal particle has been synthesized, one that is capable of exploiting energy from its environment to propel itself into a medium at tens of microns per second. These self-driven units~\cite{wang_one_2015,wang_small_2013,bechinger_active_2016,ebbens2010pursuit,palacci_photoactivated_2013,palacci_light-activated_2014, stenhammar2016light,yan2016reconfiguring} have been shown to develop unusual collective behavior and are capable of driving non-active systems they interact with in and out of equilibrium. As such, they hold promise for the development of the next generation of smart materials which can be tuned from the bottom up to acquire active properties themselves~\cite{Mallory2019Feb,Mallory2018Apr,Shaebani2020Apr,Bar2020Mar,Bechinger2016Nov,Palacci2014Nov,Chakrabarti2019}.

With this in mind, we consider a system of active colloidal particles linearly connected to form a semi-flexible filament.
Unlike other active polymer systems previously considered in the literature, where the active forces are either tangential to its backbone, mimicking the behavior of biological filaments or strings of active dipolar particles~\cite{Schaller2010Sep,Schaller2011Mar,Yan2016Jul}, or are free to rotate around each monomer, capturing the effect of active fluctuations in the fluid~\cite{Winkler2020Jul}, here we consider active forces that are aligned perpendicular to the filament backbone at every point. 


More specifically, we envision the same setup where a sequence of active particles are laterally connected to form a filament where each monomer has an axis of propulsion pointing either away from or towards a templating structure, as determined by its sequence.
One can think of this setup as a quenched, one dimensional, non-interacting Ising-like system imprinted into a semiflexible filament. Since the new generation of active particles can be driven in-or-out of equilibrium by exposure to specific light frequencies, once the monomers are activated, the particular up/down sequence along the filament, $\{s_k\}$, can now deform it and even fold it.

The overall scope of this paper is to understand whether it is possible to generate reconfigurable structures that can be consistently designed exploiting active forces. In this spirit, this model represents what is possibly the simplest setup 
to study this problem, as upon activation such filaments can be driven to collapse into complex folds, and once light is removed they re-establish an extended conformation. Crucially, our results indicate that a direct relationship between sequence and structure can be established for this system in a manner that is analogous to that existing for proteins. For this reason, we name this system  ``active colloidal folder". 
Since the metal site on active colloidal particles is quite heavy, most active colloids readily deposit at the  bottom of the solution, and perform what is effectively a two dimensional Brownian active motion with the axis of propulsion parallel to the surface that supports them. We therefore limit our study to two dimensions, consistently with the great majority of theoretical and numerical studies.


Apart from the dimensionality, the setup of our system is reminiscent of that associated to 4D printing, where folding of macroscopic filaments is driven by specific angular biases pre-encoded along a filament  backbone~\cite{Ahmed2021Jul}. {In our case, the active colloidal folders} deform by consuming free-energy in the environment to propel it active monomers. The forces generated this way result in  local and non-local pivots along the chain backbone. Figure ~\ref{model}(a) shows a simple local pivot that develops when a short segments with (say) $s=1$ is inserted in the middle of a filament of $s=-1$ monomers. The specific final angle will be a compromise between the bending rigidity of the filament, the strength of the active forces, and the overall length of the polymer.
As such, apart from trivial cases, it is not possible to, a-priory, understand the overall final conformation of the filament from simple local arguments. In this paper, we  propose a simple strategy to achieve this goal.
 
\section{Model}
\begin{figure} 
\includegraphics[width=0.35\textwidth]{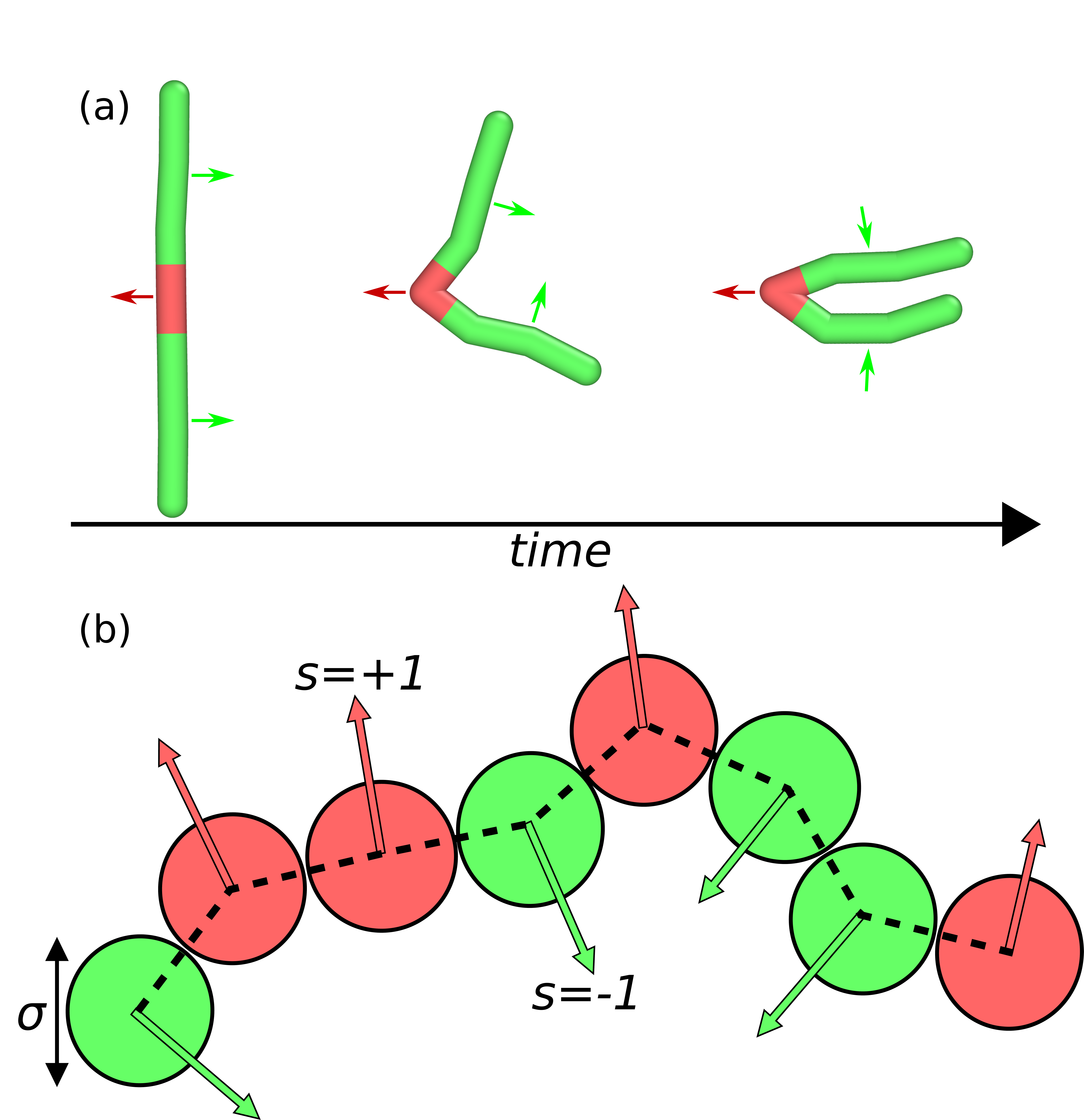}
\caption{(a) Sketch of the folding dynamics driven by active forces for a simple sequence of active forces. (b) Sketch of the model of the colloidal folder used in this work. The colors encode the direction of the forces, red for $s=+1$ and green for $s=-1$.}
\label{model}
\end{figure}

We model the active colloidal folder as a series of $N$ beads of diameter $\sigma$ linearly connected with an harmonic bond of the form $U_s(\pmb{r}_i)=k_s/2 (|\pmb{r}_i-\pmb{r}_{i+1}|-\sigma)^2$ with $k_s=800k_{\rm B}T$. The
polymer is constrained in a two-dimensional plane, and the  
energy cost associated to bending fluctuations is defined as $U_b(\pmb{t}_i)=\kappa/2\sum_{j=i-1}^{i+1} (1-\pmb{t}_i\cdot\pmb{t}_j$), where the local tangent is defined as $\pmb{t}_i=(\pmb{r}_{i+1}-\pmb{r}_{i})$, and $\kappa$ is the bending rigidity of the filament. The axis of propulsion of the active forces is maintained at all times perpendicular to the filament backbone rotated either by $+90$ degrees from the local tangent ($s_i=1$) or by $-90$ degrees 
($s_i=-1$) according to the initial sequence $\{s_k\}$ which remains fixed throughout the simulation. 
Excluded volume between any two monomers $i$ and $j$ is enforced with a truncated-shifted
Lennard-Jones potential
\begin{equation}
U(r_{ij})=4\varepsilon\left[ \left(\frac{\sigma}{r_{ij}}\right)^{12} - \left(\frac{\sigma}{r_{ij}}\right)^{6} +\frac{1}{4} \right]
\label{E1}
\end{equation}
which extends up to a cut-off value set to $2^{1/6}\sigma$, and  
$\varepsilon=1k_{\rm B}T$.
Figure~\ref{model}(b) illustrates a sketch of our model.
Each monomer undergoes Brownian dynamics at a constant temperature $T$ according to the following equations of motion: 
\begin{equation}
\label{tra} 
\frac{d\pmb{r}_i(t)}{dt} = \frac{1}{\gamma} \pmb{F}(\{r_{ij}\}) +    v_p s_i\, \pmb{n}_i + \sqrt{2D}\,\pmb{\xi}_i(t)
\end{equation}
\noindent  
where self-propulsion is introduced through a directional propelling velocity of  magnitude $v_p$ and is directed along the normal vector  to the filament at that point, $\pmb{n}_i$, with a direction defined by $s_i$. The direction of the normal vector  is computed at each monomer every time step.
The translational diffusion coefficient $D$ is related to the temperature and the translational friction $\gamma$ via the Stokes-Einstein relation $D=k_{\rm B}T\gamma^{-1}$.  The  effective solvent induced Gaussian white-noise term is characterized by $\langle \pmb{\xi}(t)\rangle = 0$ and $\langle \xi_i(t) \xi_j(t^\prime)\rangle = \delta_{ij}\delta(t-t^\prime)$. $\pmb{F}(\{r_{ij}\})$ indicates the forces arising from the inter-particle bending, bonding and excluded volume energies introduced above.
In our simulations $\sigma$ and $k_{\rm B}T$ are used as the units of length and energy scales of the system, while $\tau=\sigma^2D^{-1}$ is our unit of time. All simulations were  run with time step $\Delta t=10^{-4}\tau$, and the strength of the active forces is expressed in terms of the dimensionless P\'eclet number $Pe=v_p\sigma/D$. In our simulations, the parameters $D$, $\gamma$ and $k_BT$ are all set to 1, and the P\'eclet number is exclusively controlled by varying the active velocity $v_p$.

\section{Results \& Discussion}
\begin{figure} 
\includegraphics[width=0.45\textwidth]{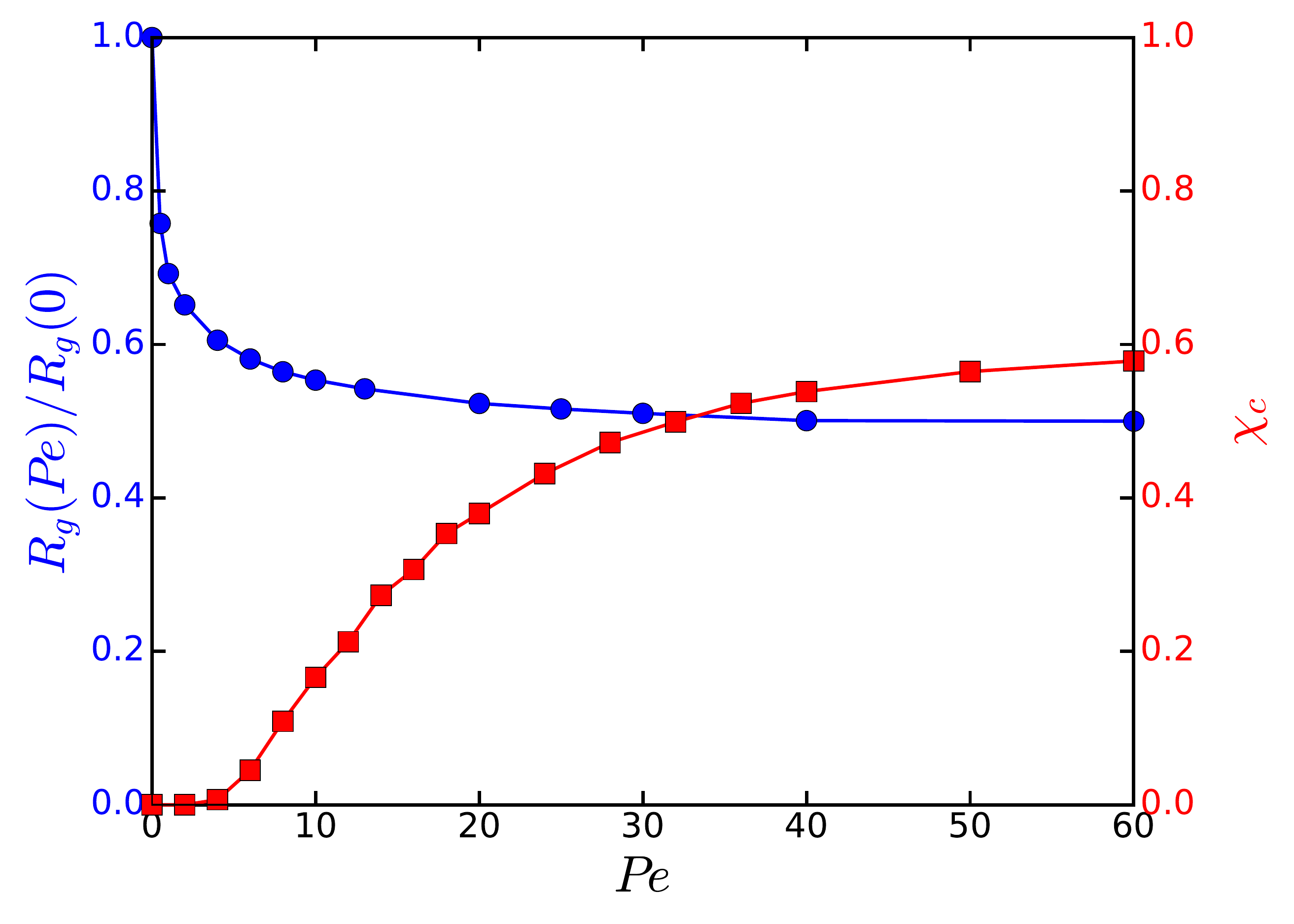}
\caption{Fraction of collapsed configurations averaged over 5000 independent sequences as a function of P\'eclet number Pe. The blue line shows the collapse in terms of the change radius of gyration of the chain with $Pe$ relative to the radius of gyration of the passive polymer. The data in red indicate how the contact number, $\chi_c$, as defined in the paper, changes as a function of $Pe$.}
\label{PD}
\end{figure}

We begin our analysis by first looking at the structural properties of the filament for a fixed bending rigidity $\kappa=10 k_{\rm B}T$ for different values of $Pe$. Specifically, we consider whether the filament collapses or remains overall extended under the action of the active forces.
For this purpose, we consider two different estimators. The first is the radius of gyration. The other is the average numbers of contacts each monomer makes with other non-bonded monomers. Here, we define the chain as collapsed if, on average, each monomer has at least 1.4 neighbors  within 1.9$\sigma$ from its center.  This selection gives a visually satisfactory representation of extended versus compact configurations for this system. Figure~\ref{PD} shows in the same plot how the radius of gyration, $R_g$, normalized by its passive ($Pe=0$) value, and  the fraction of collapsed configurations, $\chi_c$, change for different values of $Pe$. The averages are taken over a minimum of 500 independent sequences $\{s_k\}$.
In view of these results, we selected $Pe=20$ to perform an exhaustive study of the folding properties of these system at $\kappa=10k_{\rm B}T$.
For this value, the radius of gyration has already plateaued, and $\chi_c$ indicates that about 40\% of all sequences lead to a non-extended state.

\begin{figure} 
\includegraphics[width=0.475\textwidth]{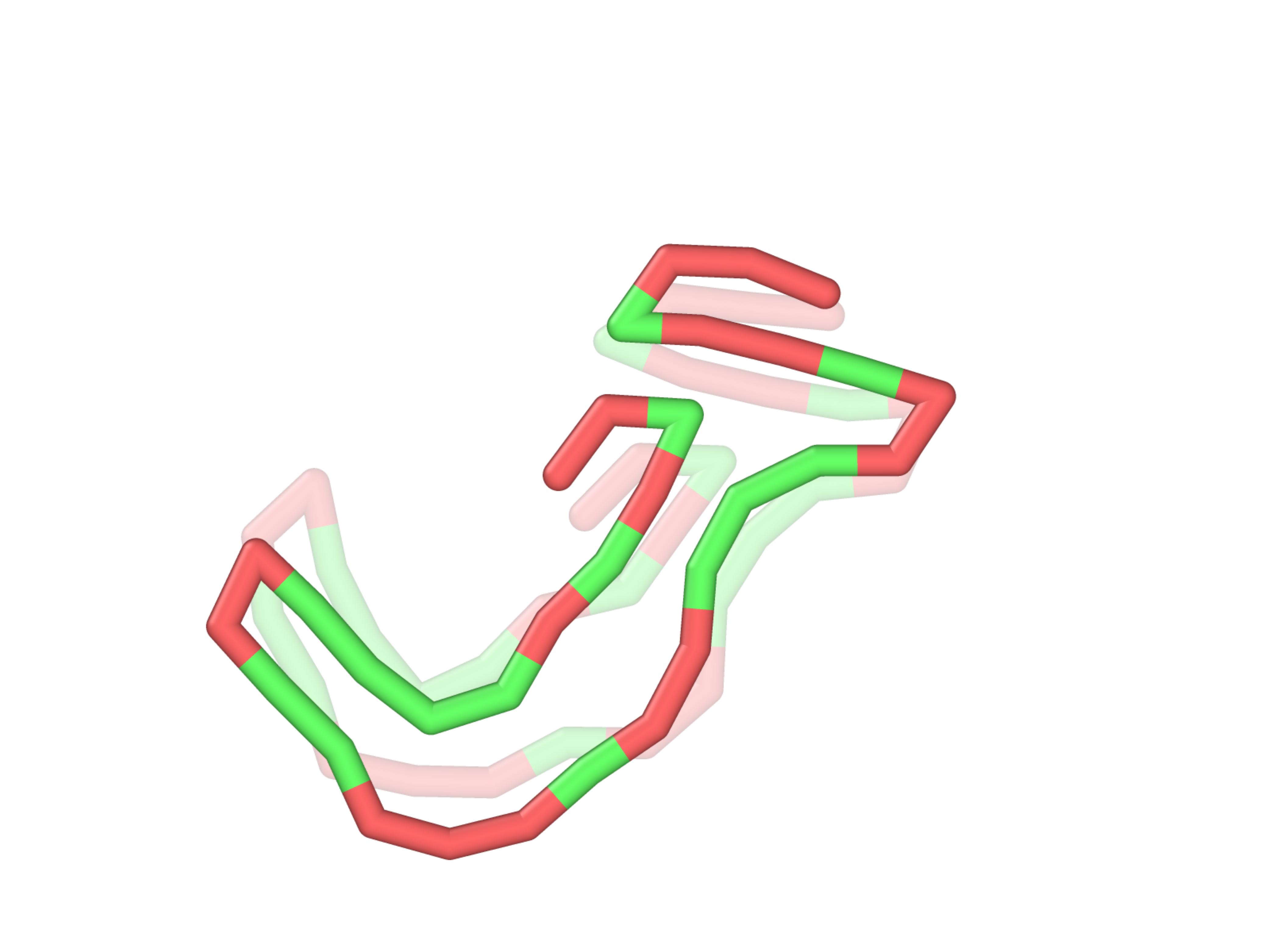}
\caption{Snapshots from our simulations highlighting the size of typical active fluctuations around a properly folded structures. The colors encode the direction of the forces, red for $s=+1$ and green for $s=-1$.}
\label{fluctuations}
\end{figure}

We should stress that although  $\chi_c$ increases with $Pe$, its asymptotic value is not expected to be equal to one, as there exist sequences $\{s_k\}$ that will maintain an extended configuration regardless of the strength of the active forces; trivially $s_i=1 \,\forall i$ or  $s_i=1$ for $i<N/2$ and $s_i=-1$ for $i\geq N/2$ lead to a translating  rods and to a rotating, rigid $S$-shaped configuration, respectively for any value of $Pe$. 
Furthermore, just because a configuration satisfies our criteria for being compact, it does not imply that it collapses into a stable structure. In fact, in most cases, the filaments are highly mobile and breath dynamically  among a range of different compact conformations.
For this reason, we rank the sequences leading to compact states depending on how long they remain into a specific conformation during our simulations and tag them as collapsed if they retain their state for at least $2\times 10^7$ iterations of our Brownian dynamics. 
The degree of state (conformational) retention is evaluated by comparing the monomers contact map at a given time to that of  a given reference state.  
A cost function 
\begin{equation}
R=\frac{1}{N(N-1)}\sum_{i>j}^{N} (d_{ij}-d^{R}_{ij})^2 \,,
\end{equation}
where $d_{ij}$ is the distance between monomers $i$ and $j$ and the suffix $R$ refers to the reference state,
is evaluated every 1000 time-steps, and whenever that is larger than a predefined value, $R_{max}=1.25\sigma$, a new reference state out of the the current configuration is created and the time counter is reset. Such a fairly large value of $R_{max}$ was selected because, unlike proteins, even the collapsed stable states in this system undergoes non-insignificant fluctuations as the filament seeks to find a balance between active, bending and thermal forces. A small angular variation at a pivot in a configuration that is not highly compact can lead to large fluctuations in the inter-particle distance down stream. Figure~\ref{fluctuations} provides a visual representation of the typical conformational deviations about a stable structure in this system.

Over a sample of $7.5\times 10^4$ independent random sequences with $N=32$ monomers, we find that about 4\% of them satisfy our criterion for a filament to collapse into a stable structure, and of this subset of sequences about 65\% systematically collapse into the same conformation akin to fast folding proteins. This was established by demanding 
that upon at least ten independent runs starting with different initial configurations, a given sequence re-folds into the same conformation. The initial configurations consisted of stretched equilibrated passive polymers, and for every new folding event we allow
each fold to re-expand (by turning off the active forces) before attempting to refold it.  So although the initial configurations are similar, a polymer with $10k_{\rm B}T$ will not be fully rigid, resulting in some variability in the shape of the extended polymer. Furthermore, the thermal fluctuations associated with the polymer dynamics will be different as different sets of random number are used for every folding event.

Given this ample sample of numerical data, we next considered whether a supervised machine learning scheme could be adopted to predict whether sequences would consistently fold into the same structure. 
To this end, we used the freely available Tensor Flow library\cite{tensorflow2015-whitepaper}. The data consisted of roughly 5,000 good folder examples and 200,000 bad folders (from which 5,000 random samples were taken). These 10,000 total samples were split (with equal bad/good ratio) into a training set ($\approx$8,000), test set ($\approx$1,000), and a validation set ($\approx$1,000). Two convolutional neural networks (CNNs), one shallow and the other deep, were trained and tested on the training and test set. The 32 binary sequence was input as a 32 node layer and the output was two nodes hot-encoded, yielding probability predictions for good or bad folding. 


Once these networks had been trained, their predictions were combined with the following rules: (1) If both agreed - use that prediction (2) Upon disagreeing - choose the one that was more confident. 
We find that the shallow and deep networks both achieve 73\% on the test data although the shallow network yields more false positives and the deep network yields more false negatives. The combined predictor produces an accuracy of 75\% on the validation set and 76\% on the remaining 195,000 bad folding sequences, indicating good transfer-ability. Comparing these numbers to a random choice success rate of 50\%, this indicates that we can use these networks to relatively accurately predict whether sequences will consistently fold into the same structure or not. 
\begin{figure} 
\includegraphics[width=0.425\textwidth]{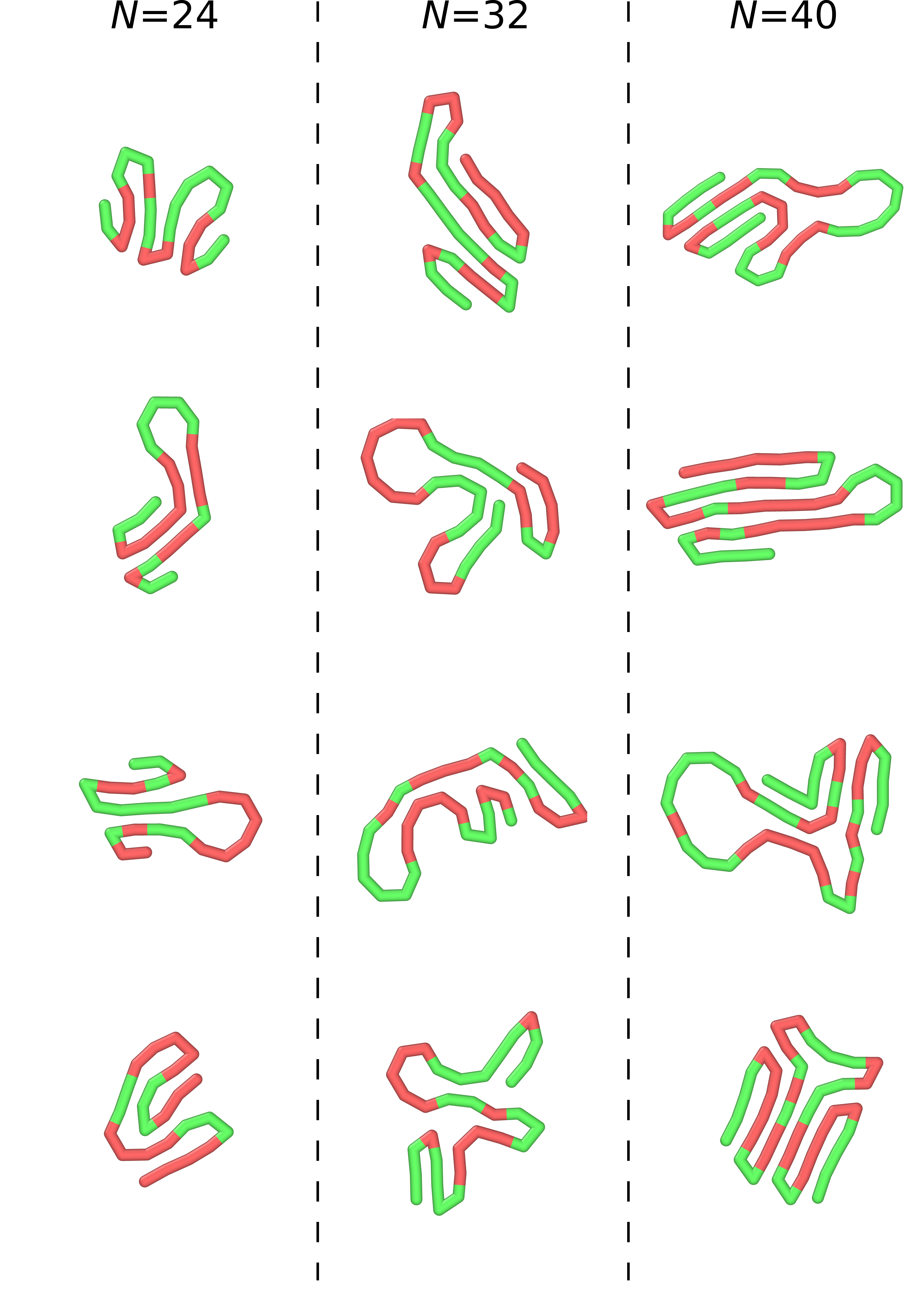}
\caption{Configurations of folded structures for chains containing $N=24, 32$ and $40$ monomers. 
These structures where initially identified as coming from good folding sequences ${s_i}^*$, and starting from a random a sequence, we used our design algorithm to recover ${s_i}^*$. The color encodes the direction of the active forces along the chain as established in Fig.~\ref{model}.} 
\label{folds}
\end{figure}

Given these results, we next attempted to design some of these compact configurations by developing a strategy to evolve an initially randomly assigned sequence along the filament into one that would consistently fold into a desired structure. Given the intrinsically non-equilibrium behavior of our system, and the peculiar nature of the driving forces resulting in its folding dynamics,  it is not obvious that our system can be mapped into an equilibrium Random Energy Model~\cite{Derrida1981Sep} . That would be a common strategy adopted in problems of protein design, for which a well established theory has been developed~\cite{Shakhnovich1998Jun} to determine the designability of a protein fold and its folding temperature. We therefore followed a different strategy. We introduced a bias potential energy function $U_f$ defined as 
\begin{equation}
U_f=\frac{b_0}{N-1}\sum_{i=1}^{N-1} \omega_i (\theta_i-\theta^{R}_i)^2 
\end{equation}
where sum runs over the  $N-1$ angles (in radians), $\theta_i$, formed by the neighboring (sequential) tangential vectors along the filament backbone. $\theta^{R}_i$ indicates what those angles would be in the target conformation we are trying to force the filament to acquire, and the $w_i$'s are a set of weights associated to each angular variable. We find that $b_0=6k_{\rm B}T$, and $\omega_i=\max(1,3|\theta^{R}_i|/\pi)$, with $\theta^{R}_i\in[-\pi,\pi]$, which gives more weight to
the more bent angles starting from $\pi/3$, makes for an efficient set of parameters for the convergence of our design scheme.
The idea is to initialize the filament with a random sequence 
$\{s_k\}$, run a short simulation in the presence of $U_f$ to obtain an estimate of $\langle U_f\rangle$, i.e.  the average energy cost required to hold the filament in the desired conformation, and finally use a simple Monte Carlo algorithm (or any other minimization scheme) to sample the space of sequences to minimize $\langle U_f\rangle$.
This scheme is analogous to that we developed for crystal design in~\cite{Miller2010Dec}.
We successfully used this procedure to obtain sequences capable of folding  filaments containing $N=24$, $N=32$ and $N=40$ monomers into four target 
structures each. Figure~\ref{folds} shows a table of images depicting the twelve target structures.

Because not all folds are designable, and unless one considers  different bending rigidities or activities per site, a randomly sketched shape is not necessarily compatible with the forces involved in our system. We therefore selected these twelve designed configurations from a set of good folding structures of known sequence, and run the design scheme as described above. Notice that not all structures that can be folded, are necessarily very compact, and unlike the case of systems at equilibrium, once the activity is removed (light is turned off) the filament reacquires an extended conformation, making these active colloidal folders a conceptually intriguing example of 
reconfigurable structures.

\section{Conclusions}

To conclude, we would like to stress that, as discussed above, several of compact structures we encountered in this study are highly dynamic and are capable of sampling a range of structures where they reside for some time before moving on to the next one, and eventually go back. It turns out that this is the behavior observed for a large number sequences, suggesting that active forces are capable of pushing the filament across different compact states that would otherwise be glassy, effectively  reducing the number of kinetic traps in the system. A similar behavior is observed in self-assembly of active particles~\cite{Mallory2018Apr}. Furthermore, one  new and exciting behavior emerging from our simulations is what we name active shapes.  These are different than the previously described dynamic structures, as some sequences lead to a smooth,  periodic motion across a set of continuously linked conformations, not unlike that observed in macroscopic actuators~\cite{Whitesides2018Apr}. We observed this behavior in some mirror-symmetric sequences at larger bending rigidities, and we are currently exploring ways of systematically extracting sequences to design active shapes.

Although in this study we gathered data for a specific bending rigidity, we expect a similar behavior to occur at larger values of $\kappa$, provided the active forces are set to be sufficiently large to overcome the local bending forces. It is also worth pointing out that we expect folding to occur also when the forces are not exactly perpendicular to the backbone of the chain
as long as the perpendicular component is the dominant one. Using a different angle between the active force and the polymer backbone ( or a distribution of angles around the perpendicular one) would still lead to folding, but the folds will be different, and since the perpendicular component would be smaller, larger active forces would be necessary for a given bending constant. 
We expect a similar trend when generalizing the model to a three-state system where some some of the monomers within the chain remain passive $s_i=0$. Folding would still occur, but active forces would need to be larger than the ones used in this study to overcome the bending energy of the polymer.

Concerning the direction of the forces, and keeping in mind a possible experimental realization of this system, it should also be pointed
out that the exact the shape of the monomers is completely immaterial for the folding to occur. So, one could in principle laterally cross-link rod-like particles, cubes with two opposite attractive faces and one of the remaining faces activated. One could also self assemble and then cross-link Janus particles with two polar hydrophobic patches and have, on a number of them, an iron-oxide element or a platinum-coated protrusion inserted between the patches.
Unless very carefully designed there could be fluctuations in the direction of the forces, but, as discussed above, some dispersion should be tolerable for
the folding to still occur.
Overall, the simple system  we presented in this paper presents a rich variety of complex behavior that has similarities to both protein folding problems and actuators. We have shown how to successfully extract sequences to target foldable structures, and have explored the efficacy of supervised machine learning algorithms in predicting such sequences. We have not attempted to use more sophisticated Artificial Intelligence techniques to improve the accuracy of the machine learning neural network, as that would be beyond the scope of this project. Work in that direction is however under way and will be published elsewhere. 

\begin{acknowledgments}
A.C. acknowledges financial support from the National Science Foundation under Grant No. DMR-2003444.
\end{acknowledgments}

\bibliography{refs}

\end{document}